\documentclass[apjl]{emulateapj}
\usepackage{graphicx}
\usepackage{amsmath, amssymb}
\usepackage{natbib}

\usepackage{color}

\begin{document}

\title{A Dynamical Potential-Density Pair for Star Clusters With Nearly Isothermal Interiors}

\author{Nicholas C. Stone$^1$ and Jeremiah P. Ostriker$^1$
\\$^1$Columbia Astrophysics Laboratory, Columbia University, New York, NY, 10027}

\begin{abstract}
We present a potential-density pair designed to model nearly isothermal star clusters (and similar self-gravitating systems) with a central core and an outer turnover radius, beyond which density falls off as $r^{-4}$.  In the intermediate zone, the profile is similar to that of an isothermal sphere (density $\rho \propto r^{-2}$), somewhat less steep than the \citet{King62} profile, and with the advantage that many dynamical quantities can be written in a simple closed form.  We derive new analytic expressions for the cluster binding energy and velocity dispersion, and apply these to create toy models for cluster core collapse and evaporation.  We fit our projected surface brightness profiles to observed globular and open clusters, and find that the quality of the fit is generally at least as good as that for the surface brightness profiles of \citet{King62}.  This model can be used for convenient computation of the dynamics and evolution of globular and nuclear star clusters.

\end{abstract}


\maketitle

\section{Introduction}

Although stellar systems can be modeled directly from observations in non-parametric ways, parametrized models are often desirable as sources of intuition, theoretical insight, or computational convenience.  Clusters of stars have been modeled with a wide variety of parametrized functions; many of these are summarized in Chapters 2 and 4 of \citet{BinTre08}.  The simplest commonly used analytic models, for example \citet{Plumme11} and \citet{Hernqu90}, are two parameter fits that do not allow for a variable ratio between outer (or half mass) radius and core size.  Since this ratio varies, and can be quite large for realistic star clusters, three parameter models generally can offer much better fits.

In this paper we use a three-parameter potential-density pair which, in projection, fits observations of globular and open clusters quite well\footnote{After this work was submitted for publication it was pointed out to us that this potential-density pair has been used for modeling gravitational lensing, e.g. \citet{KasKov93}.  Aside from the primarily numerical study of \citet{MerVal96}, which examined a triaxial analogue to this paper's potential-density pair, they have not been used for dynamical investigations of the equilibrium and evolution of nearly isothermal stellar systems.}.  Although its functional form is reminiscent of \citet{Dehnen93}, the surface density is closer to \citet{King62}, a profile widely used for fitting observed cluster surface brightnesses.  When deprojected, this model is similar to the physically motivated isothermal sphere but with the additional advantage of analytic tractability: quantities such as density, potential, and binding energy can easily be written in closed form, but cannot for the isothermal sphere or the related model of \citet{King66}.  Three important features are shared with the King profile (and the King model of \citealt{King66}): a central, constant-density core, an exterior modification that produces finite mass, and an intermediate power law zone inside the halo (or, for King, truncation) radius.  In this intermediate zone our profile scales as $\rho \propto r^{-2}$ and the King profile as $\propto r^{-3}$.  Although globular and nuclear star clusters are our primary motivation, this potential-density pair may be useful for fitting galactic potentials, as between the core and halo radii, it possesses nearly flat rotation curves.  Other possible applications include the cored dark matter profiles found in dwarf spheroidal galaxies \citep{Burker15}, or the ``$\beta$-models'' used to fit X-ray observations of galaxy clusters \citep{CavFus76}.

In \S \ref{sec:pdpair}, we present the basic formulae for our model: exactly when possible, and in approximation when necessary.  In \S \ref{sec:applications}, we consider two major applications for our model: as a theoretical testbed for studies of cluster core collapse (\S \ref{sec:cc}), and as a tool for fitting observations (\S \ref{sec:fits}).  We summarize and outline future uses for this model in \S \ref{sec:conclusions}.

\section{Potential-Density Pair}
\label{sec:pdpair}
We introduce a stellar density profile, which in functional form is related to the well-known Dehnen or ``$\eta$'' profiles \citep{Dehnen93, Tremai+94}:
\begin{equation}
\rho(r)=\frac{\rho_{\rm c}}{(1+r^2/r_{\rm c}^2)(1+r^2/r_{\rm h}^2)}. \label{eq:rho}
\end{equation}
Here $\rho_{\rm c}$ is the central density, $r_{\rm c}$ is the core radius, and $r_{\rm h}$ is the outer halo radius (which approaches the half-mass radius when $r_{\rm c} \ll r_{\rm h}$).  This density profile has been chosen to produce a flat central core contained interior to a roughly isothermal $\rho \propto r^{-2}$ cluster.  The halo densities of isolated clusters are known to fall off as $\rho \propto r^{-3.5}$ due to outer anisotropy \citep{SpiHar71, SpiSha72}; we have chosen the somewhat steeper $\rho \propto r^{-4}$ scaling for analytic tractability.  

\begin{figure}
\includegraphics[width=85mm]{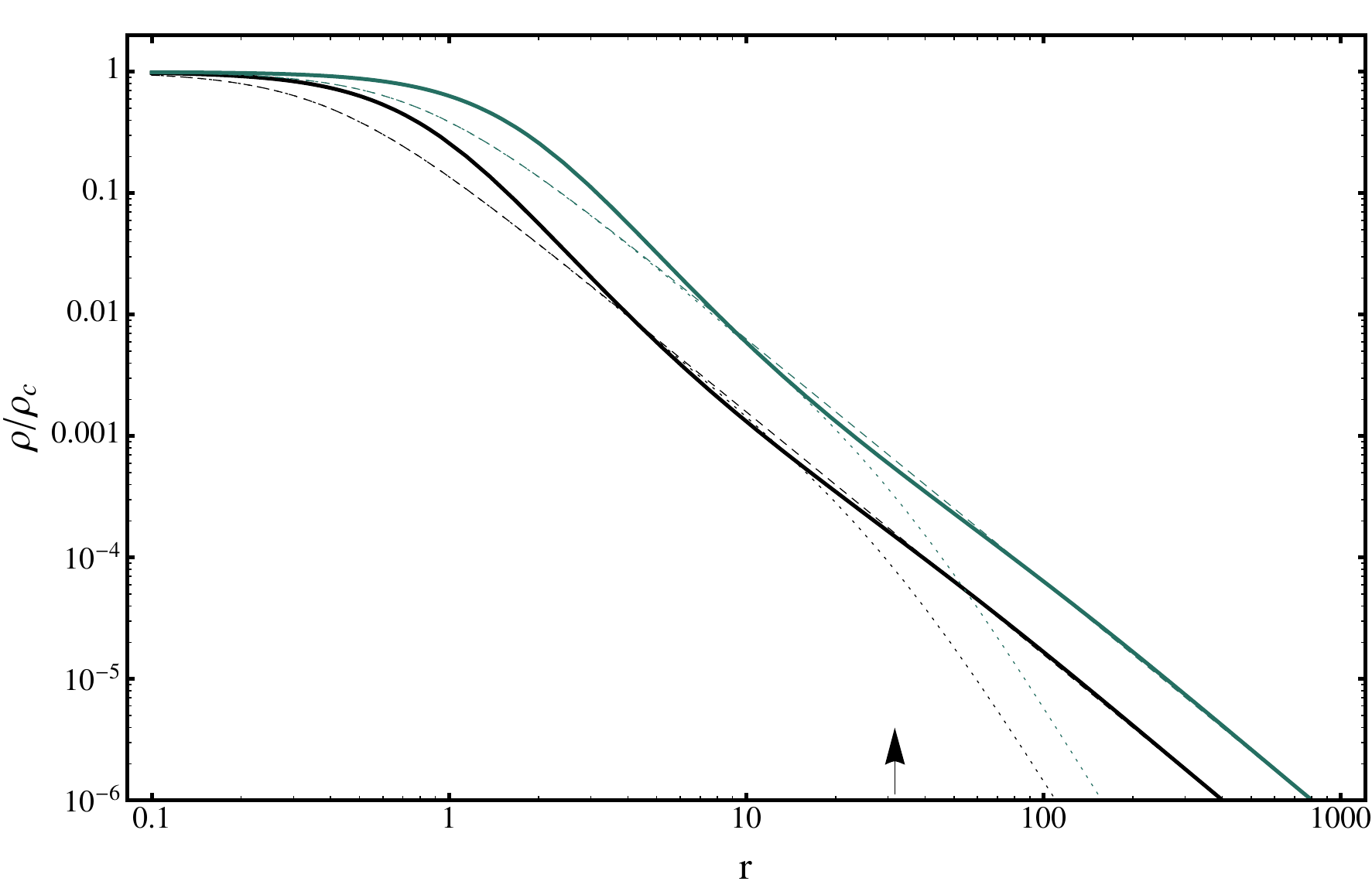}
\caption{Comparison between the classic isothermal sphere (solid lines) and Eq. \ref{eq:rho}, with halo radii at $r_{\rm h}=10^{1.5}$ (dotted curves; $r_{\rm h}$ indicated by arrow) and $r_{\rm h}=\infty$ (dashed curves).  For all curves we set $G=\rho_{\rm c}=1$, and show solutions for $\sigma=1$ (black) and $\sigma=2$ (green).  \\}
\label{isoSphere}
\end{figure}

We compare this profile to the classic isothermal sphere solution\footnote{I.e. the solution to $\frac{{\rm d}}{{\rm d}r }\left(\frac{r^2}{\rho} \frac{{\rm d}\rho}{{\rm d}r} \right)+\frac{4\pi G}{\sigma^2}r^2\rho=0$ with $\sigma$ constant and non-singular inner boundary conditions.} in Fig. \ref{isoSphere}.  Interior to the halo radius, $r_{\rm h}$, the density profiles are quite similar, provided we set the core radius $r_{\rm c}=r_{\rm iso}\sqrt{2}/3$, where the isothermal sphere core radius $r_{\rm iso}=3\sigma/\sqrt{4\pi G \rho_{\rm c}}$.  This ensures an asymptotic ($r_{\rm c} \ll r \ll r_{\rm h}$) match between $\rho(r)$ and the isothermal sphere. \\

In projection, our model's surface brightness profile is
\begin{equation}
I(R) = \frac{\Sigma_{\rm c}}{\Upsilon(r_{\rm h}-r_{\rm c})} \left(\frac{r_{\rm h}}{\sqrt{1+R^2/r_{\rm c}^2}} - \frac{r_{\rm c}}{\sqrt{1+R^2/r_{\rm h}^2}} \right), \label{eq:I}
\end{equation}
where $R$ is a projected 2D radial coordinate and $\Upsilon$ is the mass-to-light ratio (assumed to be constant in $r$).  We have defined a central surface density $\Sigma_{\rm c} = \pi \rho_{\rm c}r_{\rm c}r_{\rm h}/(r_{\rm h}+r_{\rm c})$.  For $R\lesssim r_{\rm h}$, a suitable choice of parameters can make Eq. \ref{eq:I} very similar to a King profile \citep{King62}; we elaborate on this in \S \ref{sec:fits}.  The potential is
\begin{align}
\Phi (r) =& -\frac{4\pi \rho_{\rm c}Gr_{\rm c}^2r_{\rm h}^2}{r_{\rm h}^2-r_{\rm c}^2}\bigg[\frac{r_{\rm h}}{r}\arctan(r/r_{\rm h})-\frac{r_{\rm c}}{r}\arctan(r/r_{\rm c})  \notag \\
&+\frac{1}{2}\ln\left(\frac{r^2+r_{\rm h}^2}{r^2+r_{\rm c}^2} \right) \bigg].
\end{align}
Relatedly, the mass enclosed at a radius $r$ is 
\begin{equation}
M(r)=\frac{4\pi r_{\rm c}^2r_{\rm h}^2\rho_{\rm c}}{r_{\rm h}^2-r_{\rm c}^2}\left[r_{\rm h}\arctan(r/r_{\rm h})-r_{\rm c}\arctan(r/r_{\rm c}) \right].
\end{equation}
This expression converges as $r \rightarrow \infty$, so the total cluster mass is
\begin{equation}
M_{\rm tot}=\frac{2\pi^2 r_{\rm c}^2r_{\rm h}^2\rho_{\rm c}}{r_{\rm h}+r_{\rm c}}.
\end{equation}
When $r_{\rm c} \ll r_{\rm h}$, the half-mass radius is nearly $r_{\rm h}$ and the core mass is given by
\begin{equation}
M_{\rm c} \equiv M(r_{\rm c}) \approx \frac{2(1-\pi/4)}{\pi}\frac{r_{\rm c}}{r_{\rm h}}M_{\rm tot}.
\end{equation}
The escape velocity from any point in the cluster is $v_{\rm esc}(r) = \sqrt{-2\Phi(r)}$.  At the cluster center, this becomes
\begin{equation}
v_{\rm esc}(0) = \frac{2}{\sqrt{\pi}} \sqrt{\frac{GM_{\rm tot}}{r_{\rm h}-r_{\rm c}}}\sqrt{\ln(r_{\rm h}/r_{\rm c})}.
\end{equation}

The above formulae have been used in the past to study gravitational lensing \citep[e.g. appendix A of][]{EliasD+07}; we now derive new properties of this potential-density pair more relevant for star cluster dynamics.  The total potential energy of the cluster is $W=-4\pi G \int_0^\infty r\rho(r)M(r){\rm d}r$.  This integral evaluates to
\begin{align}
W=&\frac{W_0}{r_{\rm h}-r_{\rm c}} \Big[r_{\rm c}\ln 4 + r_{\rm h}\ln 4 - 2 r_{\rm c} \ln\left(1+r_{\rm h}/r_{\rm c} \right)\label{eq:binding} \\
&  - 2 r_{\rm h} \ln\left(1+r_{\rm c}/r_{\rm h} \right) \Big],  \notag
\end{align}
where we have defined $W_0 \equiv -GM_{\rm tot}^2\pi^{-1}/(r_{\rm h}-r_{\rm c})$.  By the virial theorem, the total cluster binding energy $E=W/2$.

\begin{figure}
\includegraphics[width=85mm]{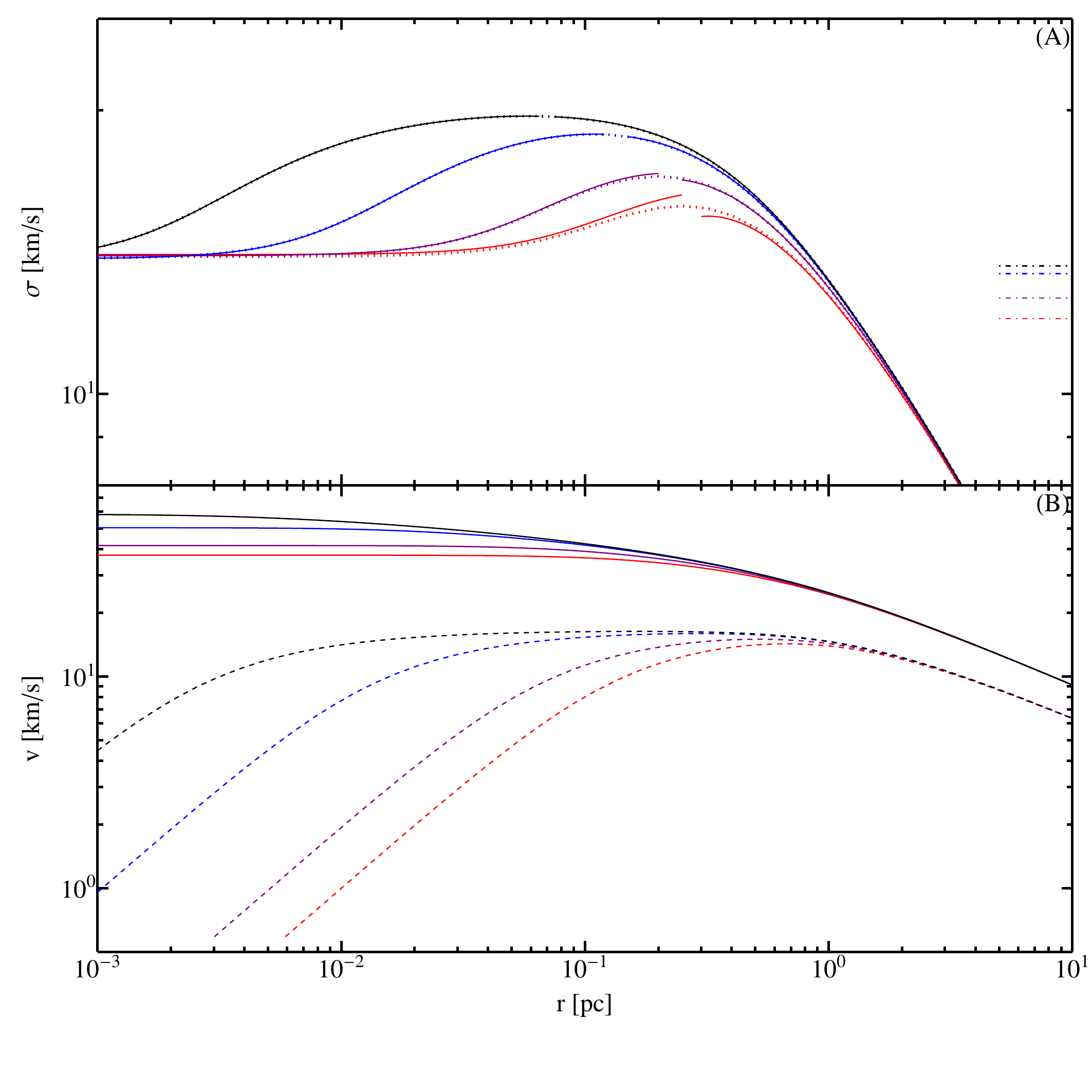}
\caption{{\it Top panel:} the velocity dispersion $\sigma$ for the potential-density pair.  Here we show four clusters with $M_{\rm tot}=10^5 M_\odot$ and $r_{\rm h}=1~{\rm pc}$; the red, purple, blue, and black lines show $r_{\rm c} = 0.1~{\rm pc}$, $r_{\rm c}=0.05~{\rm pc}$, $r_{\rm c}=0.01~{\rm pc}$, and $r_{\rm c}=0.002~{\rm pc}$.  Dot-dashed lines on the right are the virially averaged $\sigma_v$, while solid lines are approximate closed form solutions $\sigma_{\rm near}(r)$ and $\sigma_{\rm far}(r)$.  Dotted lines represent exact numerical solutions of Eq. \ref{eq:sigmaFull}.  Taken piecewise, our approximate solutions describe velocity dispersion well for $r_{\rm c}\lesssim 0.1 r_{\rm h}$, but break down severely for $r_{\rm c} \gtrsim 0.2 r_{\rm h}$.  {\it Bottom panel:} using the same line colors, dashed lines show circular velocity curves, $v_{\rm circ}(r)=\sqrt{GM(r)/r}$, and solid lines show escape velocities $v_{\rm esc}(r)=\sqrt{-2\Phi(r)}$ for each cluster. \\}
\label{JSigma}
\end{figure}

Computation of the velocity dispersion, $\sigma(r)$, is more challenging.  The virially averaged value is $\sigma_{\rm v} \equiv (-W/M_{\rm tot})^{1/2}$, but finding the radial dependence is more difficult.  Assuming isotropic velocities, the exact three-dimensional dispersion is
\begin{align}
\sigma^2(r)=& \frac{3G}{\rho(r)} \int^\infty_r \frac{M(r') \rho(r')}{r'^2}{\rm d}r' \label{eq:sigmaFull} \\
=&\frac{12\pi G\rho_{\rm c}r_{\rm c}^2r_{\rm h}^2}{r_{\rm h}^2-r_{\rm c}^2} \frac{(r^2+r_{\rm h}^2)(r^2+r_{\rm c}^2)}{r_{\rm c}^2-r_{\rm h}^2}\notag  \times \Big[Y(r, r_{\rm h}, r_{\rm c})\\
& - Y(r, r_{\rm c}, r_{\rm c})  -Y(r, r_{\rm h}, r_{\rm h}) + Y(r, r_{\rm c}, r_{\rm h})\Big] \notag
\end{align}
where
\begin{equation}
Y(r, r_1, r_2) = \int^{\infty}_{r}\frac{r_{1}}{r'^4}\frac{\arctan(r'/r_{1})}{1+r'^2/r_{2}^2}{\rm d}{r'}. \label{eq:YGeneral}
\end{equation}
Analytic approximations for $\sigma(r)$ are derived in the appendix for the limit of $r_{\rm c} \ll r_{\rm h}$.  In this regime, the central velocity dispersion is
\begin{align}
\sigma_{\rm c}^2 \equiv \lim_{r\rightarrow 0}\sigma^2 &\approx \frac{6GM_{\rm tot}(\pi^2/8-1)}{\pi r_{\rm h}} \label{eq:sigmaCentral} \\
&\approx 12\pi G \rho_{\rm c} r_{\rm c}^2 (\pi^2/8-1). \notag
\end{align}
Interestingly, in this limit, $\sigma_{\rm c}^2/\sigma_{\rm v}^2 \approx 1.011$.  In general, clusters with $r_{\rm c} \sim r_{\rm h}$ are almost perfectly isothermal for $r<r_{\rm h}/2$, while those with smaller core radii see a plateau of elevated velocity dispersion for $r_{\rm c} < r < r_{\rm h}$.  In this plateau the dispersion is nearly isothermal, with $\sigma_{\rm p}^2 \approx 3GM_{\rm tot}/(\pi r_{\rm h})$ (for $r_{\rm c} \ll r \ll r_{\rm h}$).  Although the temperature inversion visible for $r_{\rm c} \lesssim r \lesssim r_{\rm h}$ is not physical for long-term cluster equilibria, it is characteristic of a post-collapse cluster core undergoing gravothermal expansion.  Fig. \ref{JSigma} shows exact numerical solutions for $\sigma(r)$, our approximate analytic $\sigma_{\rm far}(r)$ and $\sigma_{\rm near}(r)$ derived in Appendix A, and other velocities of interest ($v_{\rm circ}$, $v_{\rm esc}$).

The isotropic distribution function (DF) can be calculated numerically using Eddington's formula, but a closed form does not appear to exist.  This is in contrast to the somewhat similar King model \citep{King66}.  It is possible, however, to modify many of the above formulae into accomodating a central black hole (BH) of mass $m_\bullet$.  Leaving $\rho(r)$ and $I(R)$ unchanged, and setting $M_\bullet(r) = M(r) + m_\bullet$ and $\Phi_\bullet(r) = \Phi(r) -Gm_\bullet/r$, we find:
\begin{align}
&W_\bullet = W - \frac{4\pi Gm_\bullet r_{\rm c}^2 r_{\rm h}^2 \rho_{\rm c} \ln (r_{\rm h}/r_{\rm c})}{r_{\rm h}^2-r_{\rm c}^2} \\
&\sigma_\bullet^{2}(r) = \sigma^2(r) + 3Gm_\bullet \left(1+\frac{r^2}{r_{\rm c}^2} \right)\left(1+\frac{r^2}{r_{\rm h}^2} \right)\Bigg[\frac{1}{r}\label{sigmaBH} \\
&-\frac{\pi(r_{\rm h}^2+r_{\rm c}r_{\rm h}+r_{\rm c}^2)}{2r_{\rm c}r_{\rm h}(r_{\rm h}+r_{\rm c})} + \frac{r_{\rm h}^2\arctan(r/r_{\rm c})}{r_{\rm c}(r_{\rm h}^2-r_{\rm c}^2)} - \frac{r_{\rm c}^2\arctan(r/r_{\rm h})}{r_{\rm h}(r_{\rm h}^2-r_{\rm c}^2)} \Bigg].\notag
\end{align}
It is not obvious, but $\sigma_\bullet^2-\sigma^2 \propto 1/r$ at large radii due to multiple high-order cancellations in Eq. \ref{sigmaBH}.

Under the approximation that $r_{\rm c} \ll r_{\rm h}$, we calculate the two-body relaxation time \citep{BinTre08},
\begin{equation}
t_{\rm r}(r)\equiv \frac{6.5\times10^{-2}~\sigma^3(r)}{G^2m_* \rho(r) \ln \Lambda},
\end{equation}
with the Coulomb logarithm $\Lambda \approx 0.4 M_{\rm tot}/m_*$ \citep{SpiHar71}.  Specifically, the central relaxation time
\begin{equation}
t_{\rm r}(0) \approx \frac{0.39}{\ln \Lambda} \sqrt{\frac{r_{\rm c}^3}{GM_{\rm tot}}} \frac{M_{\rm tot}}{m_*} \frac{\sqrt{r_{\rm c}r_{\rm h}}}{r_{\rm c}+r_{\rm h}}.
\end{equation}
Approximating the inner regions of the cluster as isothermal lets us write $t_{\rm r}(r) \approx t_{\rm r}(0)(1+r^2/r_{\rm c}^2)(1+r^2/r_{\rm h}^2)$, which is valid until the turnover in the velocity dispersion at $r\approx r_{\rm h}/2$.

\section{Applications}
\label{sec:applications}

\subsection{Single Component Cluster}
\label{sec:cc}

If we apply the potential-density pair to a cluster composed of single, identical stars of mass $m_*$, neglecting three-body and tidal capture processes, we can describe very simply the dynamics of core collapse.  In particular, under the assumption of self-similarity, we can write
\begin{equation}
\frac{{\rm d}r_{\rm c}}{{\rm d}t}\propto- \frac{r_{\rm c}}{t_{\rm r}(0)}. \label{drc}
\end{equation}
Self-similar core collapse is expected in single-mass clusters on analytic grounds \citep{LynEgg80}, and this is verified in numerical simulations \citep[and references therein]{Baumga+03}.  In the context of our model, self-similarity occurs under a simplifying assumption about heat conduction.  Specifically, we assume that the cluster lacks internal sources of heat (i.e. no binaries, no BHs, etc.) and set
\begin{equation}
\dot{E}_{\rm cond} \equiv A_{\rm cond} \frac{M(r_{\rm c})\sigma_{\rm c}^2}{2t_{\rm r}(0)} = -\frac{{\rm d}E_{\rm c}}{{\rm d}r_{\rm c}} \frac{{\rm d}r_{\rm c}}{{\rm d}t}, \label{drc2}
\end{equation}
where the binding energy of the core $E_{\rm c} \approx M_{\rm c} \sigma_{\rm c}^2/2$.  This evolution is trivially self-similar for constant density cores as long as we take the core $\sigma$ to be isothermal (a good approximation for our model), and set $A_{\rm cond}$ to a constant value.  This latter step must unfortunately be done artificially; the mild ``bump'' in our model's $\sigma$ profile prohibits a self-consistent calculation of outward heat flux.  

We find an analytic solution in the limit of $r_{\rm c}\ll r_{\rm h}$ and fixed $r_{\rm h}$:
\begin{equation}
r_{\rm c}(t) = \sqrt{r_{\rm c}^2(0)-5.1 tA_{\rm cond}\ln \Lambda \frac{m_*}{M_{\rm tot}} \sqrt{GM_{\rm tot}r_{\rm h}}  }\label{CCSS}
\end{equation} 
Thus, the core-collapse timescale in this limit is
\begin{align}
T_{\rm cc} &= \frac{1}{2A_{\rm cond}}t_{\rm r, 0}(0)  \label{eq:TCC} \\
& \approx \frac{0.19}{A_{\rm cond}\ln\Lambda} \sqrt{\frac{r_{\rm h}^3}{GM_{\rm tot}}} \frac{M_{\rm tot}}{m_*} \frac{r_{\rm c}^2}{r_{\rm h}^2}, \notag
\end{align}
where $t_{\rm r, 0}(0)$ is the initial central relaxation time.  Isotropic Fokker-Planck models take $\approx 330$ central relaxation times \citep{Cohn80} to reach total core collapse, implying $A_{\rm cond} \approx 1.5\times 10^{-3}$.  Direct N-body simulations are in reasonable agreement with this \citep{Baumga+03}.  

\begin{figure}
\includegraphics[width=85mm]{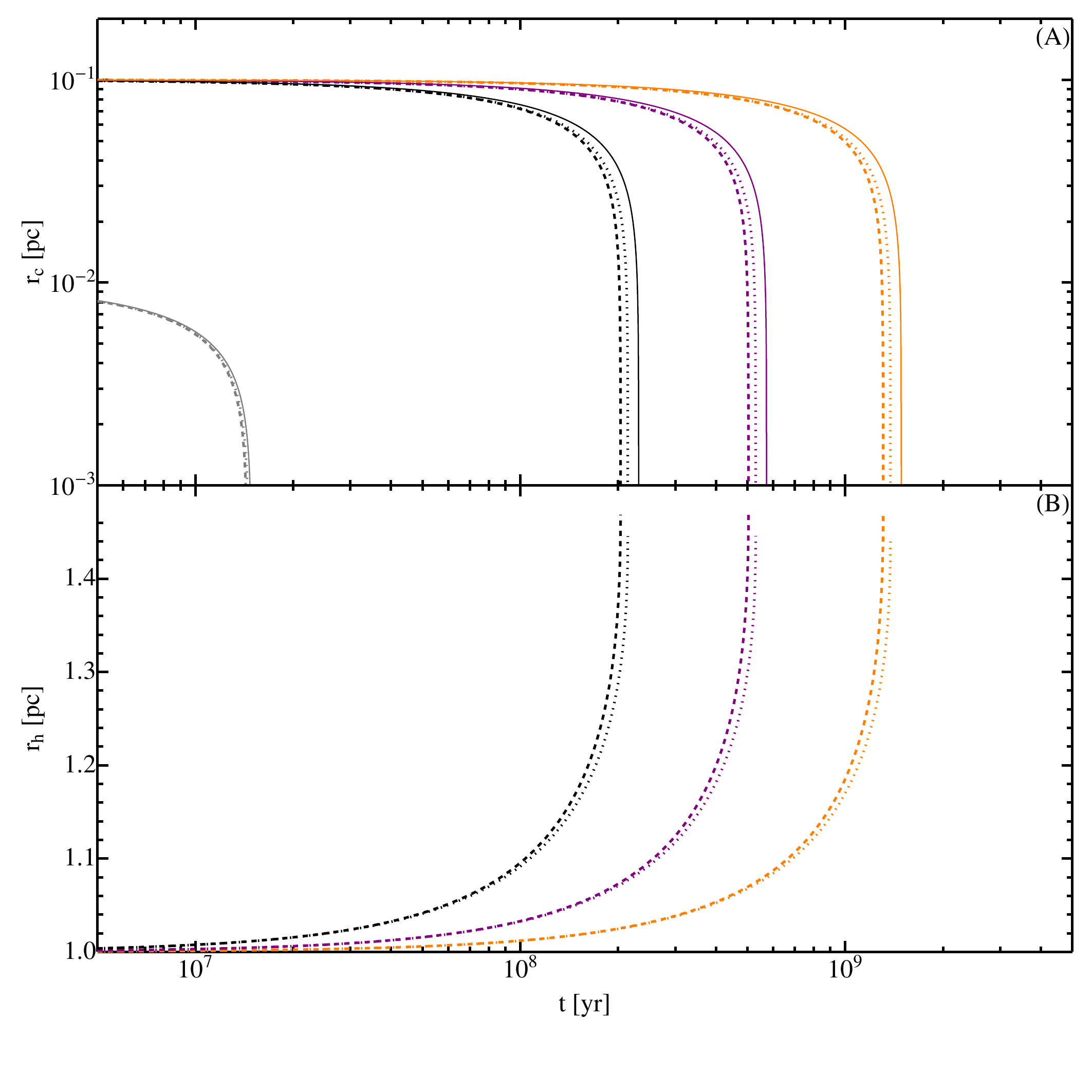}
\caption{The core radius $r_{\rm c}$ ({\it panel A}) and halo radius $r_{\rm h}$ ({\it panel B}) of an equal-mass star cluster undergoing core collapse.  The self-similar process runs away rapidly, and is highly sensitive to initial conditions.  The orange, black, and purple curves correspond to $M_{\rm tot}$ of $10^4M_\odot, 10^5 M_\odot$, and $10^6M_\odot$, respectively, and in each of these cases the initial conditions are $r_{\rm h}=1~{\rm pc}$, $r_{\rm c}=0.1~{\rm pc}$.  The grey curves are the same as the orange ones, but with $r_{\rm c}=0.01~{\rm pc}$ initially.  The thick dotted and dashed curves are numerical integrations of Eqs. \ref{drc2} and \ref{drt} with fixed and variable central $\sigma$, respectively, while the thin solid curve represents the self-similar analytic solution from Eq. \ref{CCSS}.  All solutions come to overlap as $r_{\rm c}\ll r_{\rm h}$. \\}
\label{rBoth}
\end{figure}

With this calibration, we can now self-consistently model the evolution of our cluster.  Heat flowing outward from the collapsing core will inflate the outer regions (assuming an isolated cluster unaffected by external tides) so as to satisfy ${\rm d}W/{\rm d}t=0$.  This produces the differential equation
\begin{equation}
\frac{{\rm d}r_{\rm h}}{{\rm d}t} =\frac{{\rm d}r_{\rm c}}{{\rm d}t} \frac{W- \ln\left[\frac{1}{2}(1+r_{\rm h}/r_{\rm c}) \right]W_0}{W+ \ln\left[\frac{1}{2}(1+r_{\rm c}/r_{\rm h}) \right]W_0}. \label{drt}
\end{equation}
Eqs. (\ref{drc}) and (\ref{drt}) provide a simple description of a cluster undergoing core collapse while conserving energy.  We plot the core radii and outer radii of core collapsing star clusters in Fig. \ref{rBoth}.  We see that $r_{\rm h}$ evolves very little, justifying the approximation used in Eq. \ref{CCSS}.

\begin{figure}
\includegraphics[width=85mm]{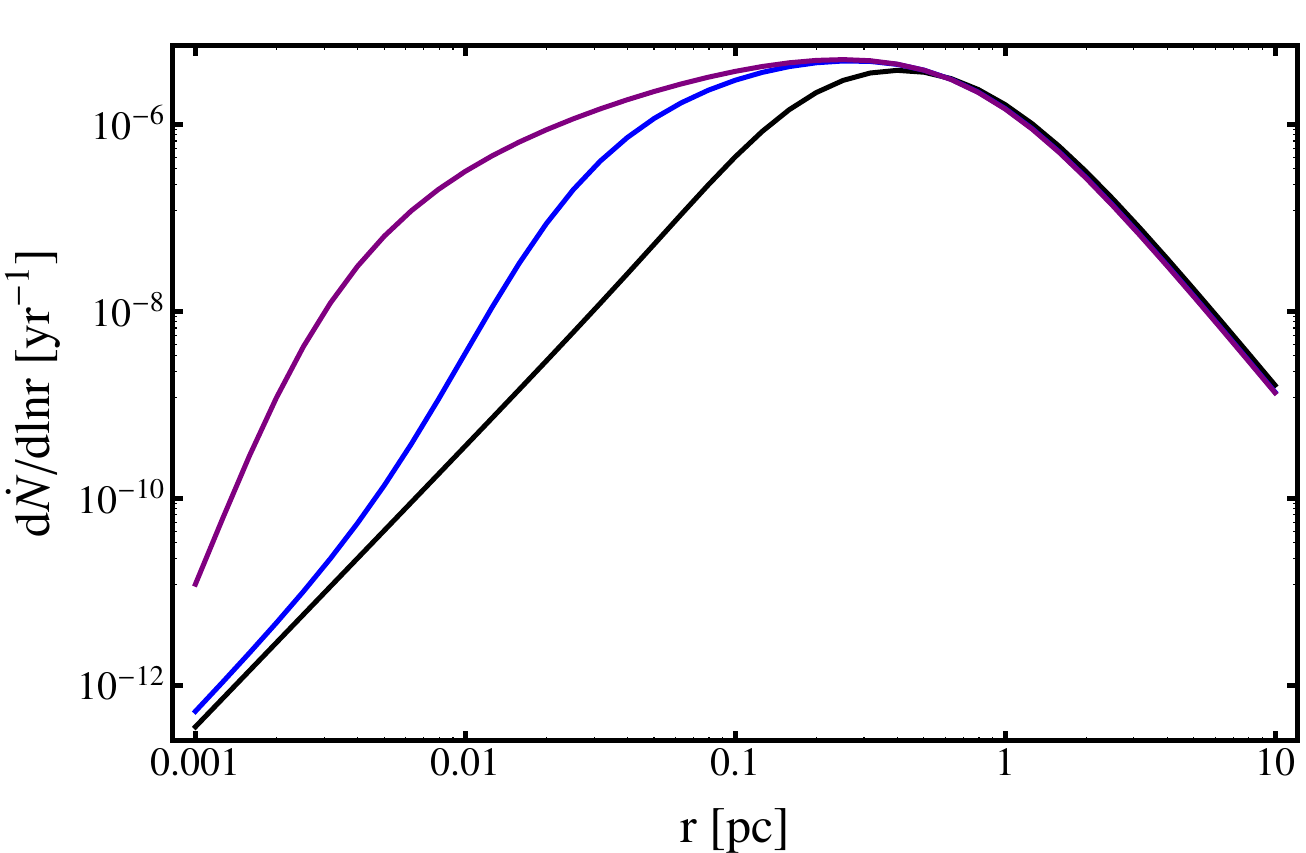}
\caption{The evaporation rate as a function of radius from clusters with $M_{\rm tot}=10^5 M_\odot$ and $r_{\rm h}=1~{\rm pc}$.  The black, blue, and purple lines correspond to $r_{\rm c}=0.1, 0.01,$ and $0.001~{\rm pc}$, respectively.  If we integrate these curves with respect to radius, we find total evaporation rates of $\dot{N}_{\rm esc} = 1.3\times 10^{-6}~{\rm yr}^{-1}, 2.2\times 10^{-6}~{\rm yr}^{-1}, $ and $2.6\times 10^{-6}~{\rm yr}^{-1}, $ respectively. \\}
\label{fig:evapRate}
\end{figure}

Competing against core collapse is the evaporation of the cluster \citep{Ambart38}, due to losses from the high-velocity tail of the velocity distribution.  We assume a Maxwellian velocity distribution $f(v) \propto \exp(-\frac{3v^2}{2\sigma^2})$, and use this to define the fraction of stars in the distribution that will escape the cluster, 
\begin{equation}
\zeta_{\rm esc}(r) = \frac{\int_{v_{\rm esc}(r)}^\infty f(v)v^2{\rm d}v}{\int_0^\infty f(v)v^2{\rm d}v} \approx \sqrt{\frac{6}{\pi}} \frac{v_{\rm esc}(r)}{\sigma(r)} {\rm e}^{-\frac{3}{2}\frac{v_{\rm esc}^2(r)}{\sigma^2(r)}}.
\end{equation}
The approximate equality assumes $v_{\rm esc}(r) \gtrsim \sigma(r)$.  Assuming that this fraction is replenished on a local relaxation time, $t_{\rm r}(r)$, the total rate of stellar evaporation from the cluster is $\dot{N}_{\rm esc} = \int^\infty_0 \frac{{\rm d}\dot{N}_{\rm esc}}{{\rm d}r}{\rm d}r$,
where 
\begin{equation}
\frac{{\rm d}\dot{N}_{\rm esc}}{{\rm d}r}{\rm d}r = 4\pi r^2 \frac{\rho(r)}{m_*} \frac{\zeta_{\rm esc}(r)}{t_{\rm r}(r)}{\rm d}r.
\end{equation}
We plot ${\rm d}\dot{N}_{\rm esc}/{\rm d}\ln r$ in Fig. \ref{fig:evapRate}.  Most of the mass flux out of the cluster comes from radii $r_{\rm c} \lesssim r \lesssim r_{\rm h}$.  For fixed $M_{\rm tot}$ and $r_{\rm h}$, decreasing the core radius $r_{\rm c}$ will marginally increase the total evaporation rate $\dot{N}_{\rm esc}$.  Unsurprisingly, the specific evaporation rate is much higher outside the core than in it.  If we approximate $\dot{N}_{\rm esc} \approx {\rm d}\dot{N}/{\rm dln}r|_{r=r_{\rm h}}$, we can compute an evaporation timescale $T_{\rm evap}=N/\dot{N}_{\rm esc}$ as
\begin{equation}
T_{\rm evap} \approx \frac{14}{\ln\Lambda} \sqrt{\frac{r_{\rm h}^3}{GM_{\rm tot}}} \frac{M_{\rm tot}}{m_*},
\end{equation}
in the limit where $r_{\rm c} \ll r_{\rm h}$.  Comparing to Eq. \ref{eq:TCC}, we see that for a cluster without internal heat sources, $T_{\rm evap}/T_{\rm cc} \approx 0.1 (r_{\rm h}/r_{\rm c})^2$: core collapse will proceed relatively unimpeded by evaporation if $r_{\rm c} \lesssim 0.1r_{\rm h}$, but significantly greater initial core radii will allow evaporation to dominate at early times.

\subsection{Parametrized Fitting to Observations}
\label{sec:fits}

\begin{figure}
\begin{tabular}{cc}
\includegraphics[width=85mm]{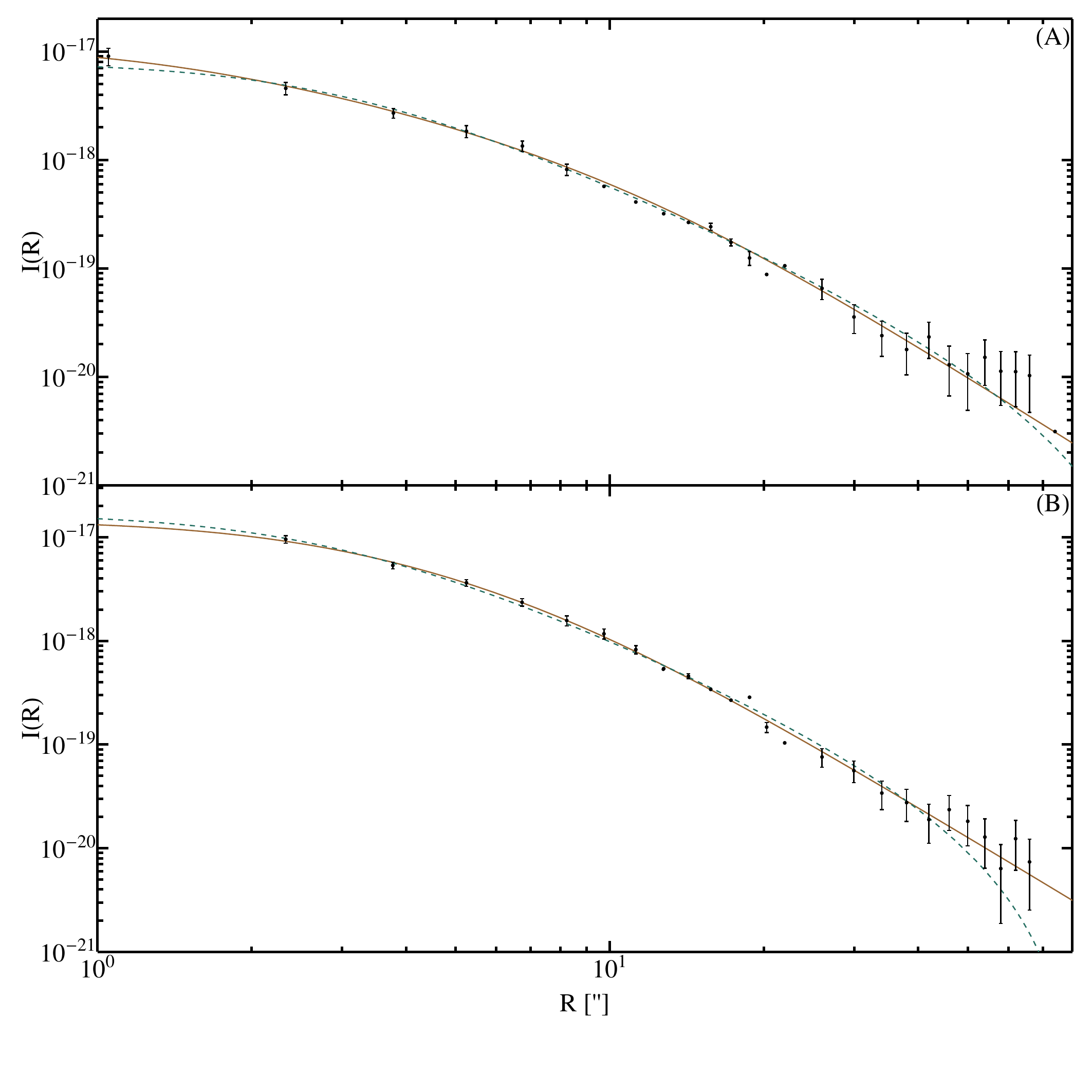}\\
\includegraphics[width=85mm]{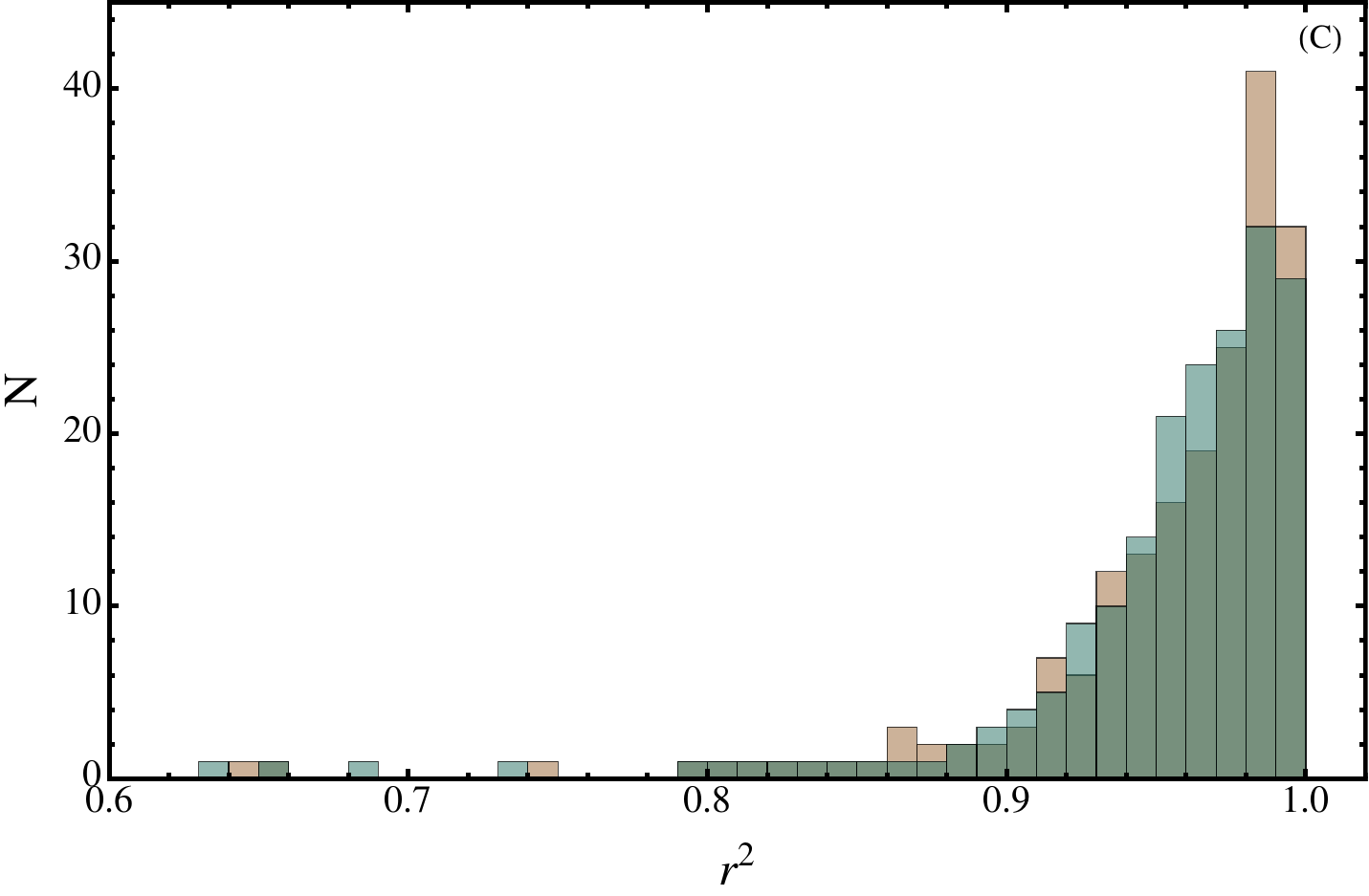}
\end{tabular}
\caption{Panels A and B show best fit parametrizations for both our model $I(R)$ (brown, solid) and the King profile $I_{\rm K}(R)$ (green, dashed).  Panel A shows the LMC globular cluster NGC 2019, where both parametrizations fit the data well.  Panel B shows the LMC globular cluster NGC 1916, where our model fits significantly better than a King profile.  In both cases data is taken from \citet{McLvan05}, and one-$\sigma$ error bars are shown (a handful of data points without error bars are shown for completeness, but were not used in the fit). Panel C shows the distribution of $r^2$ goodness-of-fit statistics for our model (brown) and the King profile (green).  The results are very similar. \\}
\label{fig:fitsBoth}
\end{figure}

Despite its pleasant analytical features, this model is primarily useful insofar as it can describe realistic star clusters.  In this subsection, we compare in greater detail to the well-known King surface brightness profile \citep{King62},
\begin{equation}
I_{\rm K}(R)= \frac{\Sigma_{\rm c, K}}{\Upsilon} \left(\frac{1}{\sqrt{1+R^2/r_{\rm c, K}^2}} - \frac{1}{\sqrt{1+r_{\rm t, K}^2/r_{\rm c, K}^2}}\right)^2.
\end{equation}
The three free parameters of the King profile are a characteristic surface density $\Sigma_{\rm c, K}$, the core radius $r_{\rm c, K}$, and the truncation radius $r_{\rm t, K}$.  The numerically computed King density profile differs from Eq. \ref{eq:rho} outside a shared constant-density core, as they possess different power law indices ($r^{-2}$ for Eq. \ref{eq:rho}, $r^{-3}$ for \citealt{King62}).  Furthermore, our $\rho(r)$ declines $\propto r^{-4}$ beyond the halo radius, while the King profile falls to $\rho=0$ at $r=r_{\rm t, K}$.  Eq. \ref{eq:rho} is likely more appropriate for clusters which are not tidally truncated, as was seen in the application of untruncated power law models to many LMC young massive clusters \citep{Elson+87}.  Applied to galactic globular clusters, these untruncated models generally outperform the King model when clusters can be resolved out to many half-mass radii, although the advantage is less clear than it is for young massive clusters \citep{McLvan05}, and the globulars can also be well-fit by the truncated (though more extended) model of \citet{Wilson75}.

We can relate Eq. \ref{eq:rho} and the King profile by equating their central surface density, core radius, and enclosed mass.  Respectively,
\begin{align}
\frac{\pi r_{\rm h}r_{\rm c} \rho_{\rm c}}{r_{\rm h}+r_{\rm c}} &= \Sigma_{\rm c, K} \left(1- \frac{1}{\sqrt{1+X}} \right)^2 \label{eq:rhoTranslation} \\
r_{\rm c} &= r_{\rm c, K} \label{eq:rcTranslation} \\
r_{\rm h}&=\frac{1}{2\pi} \Big[\frac{\sqrt{X}}{1+X}+\arctan(\sqrt{X}) \label{eq:rtTranslation} \\
& -\frac{2\ln (\sqrt{X}+\sqrt{1+X})}{\sqrt{1+X}} \Big]  \times \left(1-\frac{1}{\sqrt{1+X}} \right)^{-2}. \notag
\end{align}
In all three equations we have used $X \equiv r_{\rm t, K}^2/r_{\rm c, K}^2$.  Together, the above equations translate the parameters, $\{\Sigma_{\rm c, K}, r_{\rm c, K}, r_{\rm t, K} \}$ of a given King profile into those for our model, $\{\rho_{\rm c}, r_{\rm c}, r_{\rm h} \}$.  The reverse procedure must be done numerically because Eq. \ref{eq:rtTranslation} is nontrivially transcendental.

In Fig. \ref{fig:fitsBoth} we show best-fit parametrizations for both $I(R)$ and $I_{\rm K}(R)$.  We use the sample of 125 Milky Way globular clusters from \citet{Trager+95}, and an additional 68 LMC, SMC, and Fornax clusters from \citet[using their V-band profiles]{McLvan05}.  We fit every cluster's tabulated surface brightness profile, weighting our fits appropriately (using the measurement errors at each isophote for \citealt{McLvan05}, and more ad hoc data weights provided in \citealt{Trager+95}).  We find that within the sample, mean and median $r^2$ values for nonlinear model fitting are extremely similar for both $I(R)$ and $I_{\rm K}(R)$.  In particular, our mean $r^2$ statistic for the $I(R)$ ($I_{\rm K}(R)$) profile is 0.952 (0.942), and the corresponding median value is 0.974 (0.969).  A histogram of the $r^2$ distribution is shown in Fig. \ref{fig:fitsBoth}.

\section{Conclusions}
\label{sec:conclusions}

We have presented a novel potential-density pair that is cored, truncated, and approximately isothermal (within the halo radius).  Many important properties can be written in closed form.  In this paper, we have calculated the binding energy, surface brightness profile, central velocity dispersion, central escape velocity, and central relaxation time.  The primary quantity of interest that does not appear reducible to closed form is the stellar DF.  

The properties of this model match, surprisingly well, those of observed star clusters.  We have fit our surface brightness profile to a large number of galactic globular clusters and found generally high goodnesses of fit, comparable to or slightly better than best-fit King profiles.

Because of this model's simplicity and analytic tractability, it is well-suited for the idealized study of cluster dynamics.  In this paper we have presented a closed-form description of the evaporation and core collapse of single-component, single-particle clusters, and we hope to apply this model to other dynamical processes in the future.  

Finally, we note that the cluster model introduced here can facilitate inexpensive computation of the dynamics of mass segregated subclusters.  In particular, this potential-density pair can provide a dynamic background potential in which to run N-body simulations of the central subcluster.  We are currently using this technique to investigate intermediate mass black hole formation due to collisional runaways in nuclear star clusters \citep{Kuepper+15}, but it could also be applied to open or globular clusters.

\section*{Acknowledgments}
We would thank Haldan Cohn, Bence Kocsis, Andreas K{\"u}pper, Nathan Leigh, and Scott Tremaine for helpful discussions.  We also thank Jens Hjorth for pointing out the use of this potential-density pair in the strong lensing community.  N.S. acknowledges support through NSF grant AST-1410950 and from the Alfred P. Sloan Foundation to Brian Metzger.  This work was aided by the hospitality of the Aspen Center for Physics.

\appendix
\section{Velocity Dispersion}
Here we derive the approximate formulas for the velocity dispersion near ($\sigma_{\rm near}$; $r \ll r_{\rm h}$) and far ($\sigma_{\rm far}$; $r \gg r_{\rm c}$) from the center of the cluster.  The integrals $Y(r, r_{\rm c}, r_{\rm c})$ and $Y(r, r_{\rm h}, r_{\rm h})$ from Eq. \eqref{eq:sigmaFull} evaluate to elementary functions:
\begin{equation}
Y(r, r_1, r_1)= \frac{-1}{24r^3 r_1^2} \Bigg[4\pi r_1^3 +  4r r_1^2 - 12\pi r^2 r_1 + 3\pi^2r^3 + \left(24r^2r_1 - 8r_1^3 \right)\arctan\frac{r_1}{r}-16r^3 \ln \left(1+ \frac{r_1^2}{r^2} \right) - 12r^3\arctan^2\frac{r}{r_1} \Bigg].  \label{eq:Y23}
\end{equation}
$Y(r, r_{\rm h}, r_{\rm c})$ and $Y(r, r_{\rm c}, r_{\rm h})$ are dilogarithms, but we can find approximate solutions by Taylor expanding the arctan terms in the integrands of $Y(r, r_{\rm h}, r_{\rm c})$ and $Y(r, r_{\rm c}, r_{\rm h})$ in the limit $r' \ll r_{\rm h}$.  These integrands are dominated by contributions from $r' \approx r_{\rm c}$, so this expansion is valid when $r_{\rm c} \ll r_{\rm h}$.  These approximate solutions are 
\begin{align}
Y(r, r_{\rm h}, r_{\rm c})& \approx \frac{1}{2r^2} + \left(\frac{1}{6r_{\rm h}^2} + \frac{1}{2r_{\rm c}^2} \right)\ln\left(\frac{r^2}{r^2+r_{\rm c}^2} \right) \label{eq:Y1} \\
Y(r, r_{\rm c}, r_{\rm h})& \approx  \frac{1}{6r^3}\Bigg[\pi r_{\rm c} + r - \frac{3\pi r^2 r_{\rm c}}{r_{\rm h}^2} + 2r_{\rm c} \arctan\frac{r_{\rm c}}{r}\left(\frac{3r^2}{r_{\rm h}^2} -1\right) \Bigg] -\left(\frac{1}{r_{\rm c}^2} + \frac{3}{r_{\rm h}^2} \right) \ln \left(1+\frac{r_{\rm c}^2}{r^2} \right). \label{eq:Y4} 
\end{align}
Eqs. (\ref{eq:sigmaFull}) and (\ref{eq:Y23}-\ref{eq:Y4}) define $\sigma_{\rm near}(r)$, the approximate $\sigma(r)$ when $r \ll r_{\rm h}$ and $r_{\rm c} \ll r_{\rm h}$.  We likewise derive a $\sigma_{\rm far}(r)$ valid for $r \gg r_{\rm c}$ and $r_{\rm c} \ll r_{\rm h}$ by making the approximation that $r \gg r_{\rm c}$ in Eq. \eqref{eq:sigmaFull}.  Specifically,
\begin{align}
\sigma^2_{\rm far}(r)&=\frac{6GM_{\rm tot}(r^2+r_{\rm h}^2)(r^2+r_{\rm c}^2)}{\pi r_{\rm h}} \Bigg[ \frac{\pi^2}{8} + \frac{\pi}{6r^3r_{\rm h}}+\frac{1}{r^2r_{\rm h}^2} - \frac{\pi}{2rr_{\rm h}^3} - \frac{\arctan^2\frac{r}{r_{\rm h}}}{2r_{\rm h}^4}- \frac{\arctan\frac{r_{\rm h}}{r}}{3r^2r_{\rm h}} \\
&+ \frac{\arctan\frac{r_{\rm h}}{r}}{rr_{\rm h}^3} - \frac{2\ln \left(1+r_{\rm h}^2/r^2 \right)}{3r_{\rm h}^4} \Bigg].\notag
\end{align}
The limit is nontrivial, but cancellations of first through fourth order terms give $\sigma_{\rm far}^2 \approx 3GM_{\rm tot}/(5r)$ when $r\gg r_{\rm h}$.  When $r_{\rm c} \ll r_{\rm h}$, $\sigma_{\rm near}(r)$ and $\sigma_{\rm far}(r)$ can be joined piecewise to approximate $\sigma(r)$ for all $r$.

\end{document}